\documentclass[aps,prl,superscriptaddress,groupedaddress,reprint]{revtex4-1}

\usepackage{graphicx,dcolumn,bm,amssymb,tipa,textcomp,epstopdf}

\newcommand{\bra}[1]{\left\langle #1\right|}
\newcommand{\ket}[1]{\left| #1\right\rangle}

\newcommand{\ketbra}[2]{\left|#1\right\rangle\left\langle#2\right|}

\newcommand{\iea}[0]{{\it et al.~}}
\newcommand{\ieac}[0]{{\it et al., }}
\newcommand{\adag}[0]{$\hat a^\dagger$ }
\newcommand{\hata}[0]{$\hat a$ }
\newcommand{\pr}[0]{{\rm pr}}
\newcommand{\tr}[0]{{\rm Tr}}

\newcommand{\eqref}[1]{(\ref{#1})}
\newcommand{\eeqref}[1]{Eq.~(\ref{#1})}

\begin{document}

\title{Experimental characterization of bosonic photon creation and annihilation operators}

\author{R.~Kumar}\affiliation{Institute for Quantum Information Science, University of Calgary, Calgary, Canada, T2N 1N4}

\author{E.~Barrios}\affiliation{Institute for Quantum Information Science, University of Calgary, Calgary, Canada, T2N 1N4}

\author{C.~Kupchak}\affiliation{Institute for Quantum Information Science, University of Calgary, Calgary, Canada, T2N 1N4}

\author{A.~I.~Lvovsky}\affiliation{Institute for Quantum Information Science, University of Calgary, Calgary, Canada, T2N 1N4}\affiliation{Russian Quantum Center, 100 Novaya St., Skolkovo,
Moscow region, 143025, Russia}
\email{lvov@ucalgary.ca}

\begin{abstract}
The photon creation and annihilation operators are cornerstones of the quantum description of the electromagnetic field. They signify the isomorphism of the optical Hilbert space to that of the harmonic oscillator and the bosonic nature of photons. We perform complete experimental characterization (quantum process tomography) of these operators. By measuring their effect on coherent states, we obtain their process tensor in the Fock basis, which explicitly shows the ``raising" and ``lowering" properties of these operators with respect to photon number states. This is the first experimental demonstration of complete tomography of non-deterministic quantum processes.
\end{abstract}
\date{\today}


\maketitle

Quantum operators of annihilation \hata and creation \adag of bosonic particles act on states with a definite number $m$ of identical particles, resulting in that number being incremented or decremented, respectively: \begin{eqnarray}\label{an}
\hat a^\dag\ket m &=& \sqrt{m+1}\ket{m+1} \\ \nonumber
\hat a\ket m &=& \sqrt m\ket{m-1}.
\end{eqnarray}
First proposed by Dirac in his 1927 \emph{Quantum Theory of the Emission and Absorption of Radiation} \cite{dirac27}, these operators play an enormous role in many fields of physics and chemistry: quantum mechanics, quantum optics, quantum chemistry, quantum field theory and condensed matter physics. Specializing to optics, they are instrumental in quantum description of light, giving rise to many fundamental phenomena such as spontaneous emission, Lamb shift, Casimir force and lasing. Equally important is practical implementation of photon creation and annihilation, which provides us with a universal toolbox for manufacturing arbitrary quantum states of light, required for quantum information processing and quantum communications \cite{barb10,cerf}.

Implementation of \hata and \adag is however challenging. This is because these operators do not preserve the trace of a state's density matrix, which means they cannot occur in the framework of deterministic Hamiltonian evolution of a physical system. 
Therefore bosonic creation and annihilation can be obtained in the laboratory only in an approximate,  non-deterministic fashion. That is, the action of operators occurs with probability less than one, but is heralded by a classical event. 

First successful realization of the optical photon creation operator in this manner in 2004 \cite{zav04} gave rise to a new class of states known as photon added states. In 2006, Ourjoumtsev and colleagues applied the photon annihilation operation to the squeezed states generating optical ``Schr\"odinger cats" \cite{gar06}. Neergaard-Nielsen \iea generalized this approach in 2010 to generating arbitrary continuous-variable qubits \cite{nie10}. In 2007, Parigi \iea verified non-commutativity of \hata and \adag in application to the thermal state \cite{par07}. The photon annihilation operator has been used for continuous-variable entanglement distillation in 2010 by Takahashi and co-workers \cite{tak10}. Experimental recording of photon creation and annihilation events in the time domain has been reported by Gleyzes \iea \cite{haroche07}.

In order to include the photon creation and annihilation operations into the quantum technology toolbox, we need to develop methods of their characterization and performance testing. This is the purpose of the present work. We implement  \hata and \adag experimentally and analyze them as quantum ``black boxes", or quantum processes. By probing them with coherent optical states (weak laser pulses) of different amplitudes and measuring the quantum fluctuations of the output electromagnetic field, we determine how these black boxes would affect any arbitrary state of light within a practically relevant subspace of the optical Hilbert space. As a result, for the first time since the photon creation and annihilation operators have been discovered, we explicitly observe their action on the photon number states to be in accordance with \eeqref{an}.

The method we employ for the characterization of quantum processes relies on the optical equivalence theorem. According to that theorem, the density operator $\hat\rho$ of an arbitrary state can be written as a linear combination of coherent-state density operators,
\begin{equation}
 \hat\rho = \int P_{\hat\rho}(\alpha) \ket{\alpha}\bra{\alpha} d^2\alpha,
\end{equation}
where $P_{\hat\rho}(\alpha)$ is the Glauber-Sudarshan P function of state $\hat\rho$. Further, since  every quantum process $\mathcal{E}$ (in this case, photon creation and annihilation) is a linear map with respect to density matrices, we can write the process output as
\begin{equation}\label{Erho}
\mathcal{E}(\hat\rho) = \int P_{\hat\rho} \mathcal{E}(\ket{\alpha}\bra{\alpha}) d^2\alpha.
\end{equation}
If we know $\mathcal{E}(\ket{\alpha}\bra{\alpha})$ for every coherent state $\ket\alpha$, we can determine the process output $\mathcal{E}(\hat\rho)$ for any state $\hat\rho$.

This is useful because optical states that are employed in quantum information processing (for example, number states or their superpositions) are highly nonclassical and cannot be generated easily. In contrast, coherent states are directly obtained from lasers. Our method permits us, by probing the ``black box" with simple laser pulses, to learn its effect on any other state of light, however complicated it may be. In the past, this approach, referred to as coherent-state quantum process tomography (csQPT) \cite{lob08,sal10,anis11}, has been applied to the processes of attenuation, phase shift \cite{lob08} and quantum optical memory \cite{lob09}. A closely related method has been used for the quantum characterization of optical detectors \cite{walmsley08,walmsley12}

The result of csQPT --- the data about the process --- can be compactly written in the form of a \emph{process tensor}. This is a rank-4 tensor $\mathcal{E}_{jk}^{mn}$ such that, for any process input $\hat\rho$, the density operator of the process output in the photon number basis is given by $\left[ \mathcal{E}(\hat\rho) \right]_{jk} = \sum_{m,n} \mathcal{E}_{jk}^{mn} \hat\rho_{mn}$. The process tensor is calculated according to
\begin{equation}
  \mathcal{E}^{mn}_{jk}=\int  P_{mn}(\alpha)
  \bra j
\mathcal{E}(\ket{\alpha}\bra{\alpha})
\ket k\mathrm{d}^{2}\alpha,
\label{sup}
\end{equation}
where $P_{mn}(\alpha)$ is the P function of operator $\ketbra m n$. Computation of the process tensor is complicated by highly singular nature of this function; Refs~\cite{lob08,sal10,anis11,barb12} elaborate different ways of resolving this complication. Another practical issue is associated with the infinite dimension of the optical  Hilbert space.  In csQPT experiments, the process tensor is evaluated for a subspace  $\mathcal{H}(n_{\rm max})$ spanned by number states up to a certain cut-off value, $n_{\rm max}$. The choice of $n_{\rm max}$ is determined by the maximum amplitude  $\alpha_{\rm max}$ of the set of coherent probe states used in the experiment, as well as the reconstruction method used. In our work, $n_{\rm max}=7$.

\begin{figure}[h]
\includegraphics[width=\columnwidth]{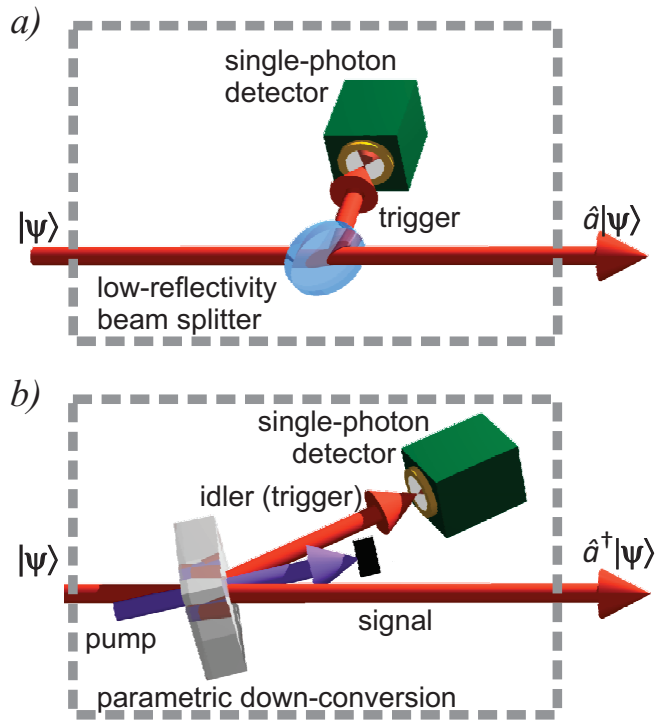}
\caption{Experimental setups for photon annihilation (a) and creation (b). The processes are heralded by ``clicks" in single-photon detectors.  }
\label{fig:setup}
\end{figure}

Practical realization of photon annihilation employs a low reflectivity beam splitter, through which the target state $\ket\psi$ is transmitted [Fig.~\ref{fig:setup}(a)]. Detection of a single photon in the reflection channel indicates that this photon has been removed from state $\ket\psi$. In this case, the state emerging in the transmission channel of the beam splitter is approximated by $\hat a\ket\psi$.

For photon creation, low-amplitude spontaneous parametric down-conversion (SPDC) in a nonlinear optical crystal can be employed. The target state enters the signal SPDC mode from the back of the crystal [Fig.~\ref{fig:setup}(b)]. If SPDC occurs, operators \adag act simultaneously on both the signal and idler modes. If this event is heralded by the photon detector in the idler channel, the target state becomes $\hat a^\dagger\ket\psi$ \cite{suppInfo}.

In the experiment, we employ a mode-locked Ti:Sapphire laser (Coherent Mira 900), which emits transform-limited pulses at $\sim$790 nm with a repetition rate of 76 MHz and a pulse width of $\sim$1.6 ps. Heralding photons are registered by PerkinElmer SPCM-AQR-14-FC single-photon detectors. The field quadratures of output states are measured by means of high-bandwidth balanced homodyne detectors \cite{leobook,lvovrmp,kumar11}. Both the probe field and the local oscillator field for homodyne detection are obtained from the master laser. The amplitude of the probe field is varied using a half-wave plate and a polarizing beam splitter.

In order to obtain SPDC, required for the photon creation, the light from the master laser is frequency doubled in a single pass through a 17-mm long lithium triborate (LBO) crystal, yielding a typical $\sim$80 mW average second-harmonic power after spatial filtering. This field is focused, with a waist of 100 $\mu$m, into a 2-mm long periodically poled potassisum-titanyl phosphate crystal. This crystal is phase-matched for type II SPDC, with the signal and idler modes being spatially and spectrally degenerate but having orthogonal polarizations.

\begin{figure}[h]
\includegraphics[width=0.7\columnwidth]{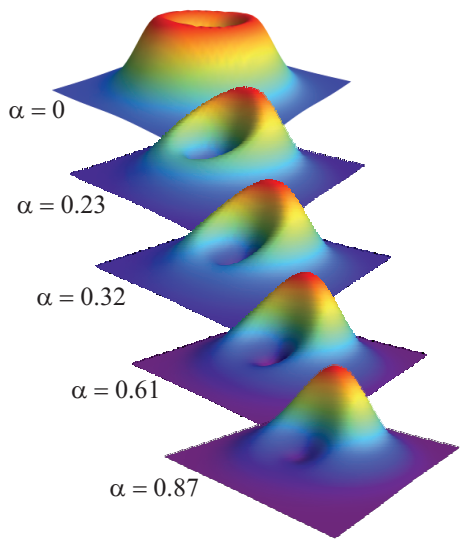}
\caption{ The Wigner functions of coherent states of different amplitudes subjected to the photon creation operator. }
\label{fig:Wigners}
\end{figure}

We acquire a set of field quadrature data for the outputs of both processes for a set of probe coherent states with amplitudes ranging from $0$ to about $1.7$. For the photon annihilation process, the output states are almost identical to the input states, as they only undergo slight attenuation propagating through the beam splitter [Fig.~\ref{fig:setup}(a)]. This is to be expected, because coherent states are eigenstates of \hata. Photon creation, on the other hand, strongly changes the nature of the state, producing single-photon added coherent states, studied in detail theoretically by Agarwal and Tara \cite{aga91}, experimentally by Zavatta \iea \cite{zav04,par07}. Highly nonclassical nature of these states is evidenced by their Wigner functions (Fig.~\ref{fig:Wigners}), which are explicitly non-Gaussian and negative in some regions of the phase space. 

The processes we study are non-deterministic, and their probability of occurrence depends on the input state. Accounting for this dependence is crucial for the correct reconstruction. In csQPT, this is done by renormalizing the process output for the probe states so that $\tr[\mathcal{E}(\ket{\alpha}\bra{\alpha})]$ in Eqs.~\eqref{Erho} and \eqref{sup} is proportional to the probability of the heralding event \cite{sal10}. To illustrate the significance of this step, it is instructive to apply \eeqref{Erho} to the photon annihilation operator, such that $\mathcal{E}(\ketbra \alpha \alpha )=|\alpha|^2 \ketbra \alpha \alpha$. If the coefficient $|\alpha|^2$, responsible for the non-deterministic nature of $\hat a$, is neglected, we would obtain the identity process. We see, remarkably, that the ``lowering" feature of \hata arises in csQPT entirely due to the variation of the event probability as a function of the probe amplitude, rather than transformation of the probe state itself.

\begin{figure}[h]
\includegraphics[width=\columnwidth]{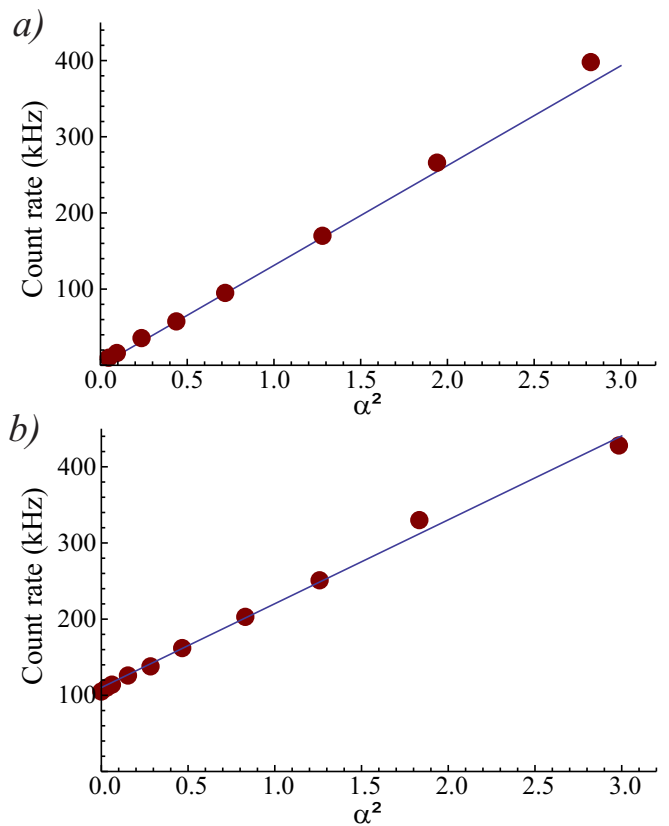}
\caption{Event count rate for the photon annihilation (a) and creation (b) operators as a function of the input coherent state amplitudes in the modes to which these operators are applied \cite{suppInfo}. The solid lines show the expected theoretical dependencies \eqref{probs}, with the vertical scale fit to the data. }
\label{fig:countrates}
\end{figure}

The information on the heralding event probability is obtained by keeping track of the photon count rates for various input states. Theoretically, we expect these rates to behave as
\begin{equation}\label{probs}
\pr_{\hat a}(\alpha)\propto\bra\alpha\hat a^\dag\hat a\ket\alpha=\alpha^2;\ \pr_{\hat a^\dag}(\alpha)\propto\bra\alpha\hat a\hat a^\dag\ket\alpha=1+\alpha^2.
\end{equation}
The experimentally observed dependencies are consistent with these expectations as displayed in Fig.~\ref{fig:countrates}. 

We use the iterative algorithm \cite{anis11} to reconstruct the process tensors directly from the acquired field quadrature data. The algorithm makes use of the Jamiolkowski isomorphism between quantum processes applied to Hilbert space $\mathcal{H}$ and positive semidefinite operators over the Hilbert space $\mathcal{H}\otimes \mathcal{H}$ \cite{hradil_book}. In this way, the task of process reconstruction is reduced to the known problem of state reconstruction \cite{lvov04,knill07,lvovrmp}. This scheme guarantees that the resulting process is physically consistent, i.e. completely positive. Furthermore, it permits us to incorporate correction for experimental imperfections into the reconstruction procedure \cite{suppInfo}.

In order to account for the non-deterministic nature of the processes being reconstructed, we introduce an additional, fictitious state $\ket{\emptyset}$ into the Hilbert space. The process can then be treated as deterministic: events in which no ``click" occurred are interpreted as events in which the process has generated state $\ket{\emptyset}$ in the output \cite{hradil_book, sal10}.



\begin{figure}[h]
\includegraphics[width=\columnwidth]{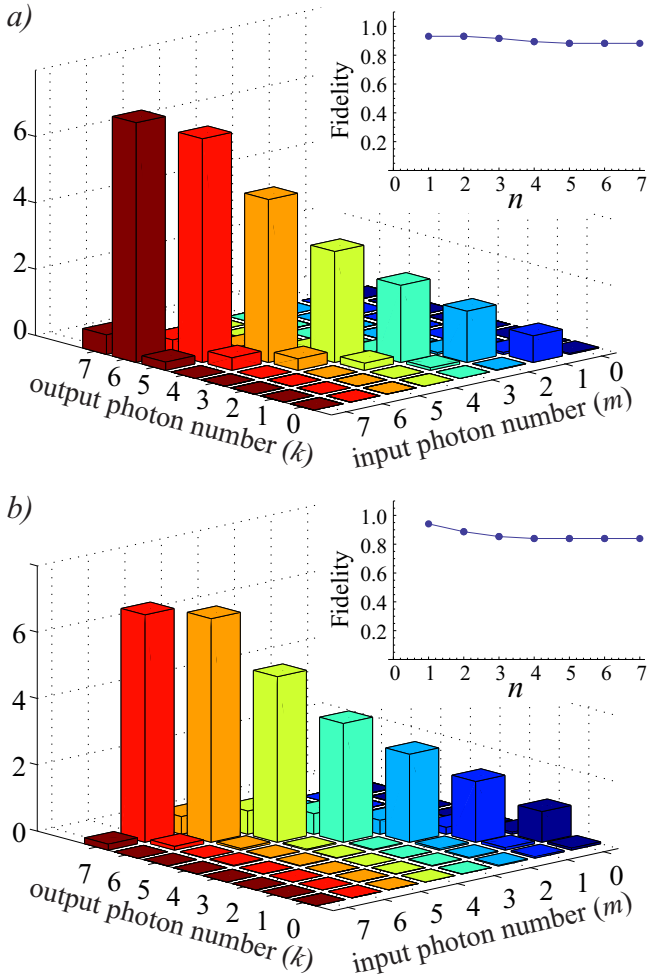}
\caption{ The ``diagonal" values of the process tensor {\Large$\varepsilon$}$_{kk}^{mm}$ of the photon annihilation (a) and and creation (b) reconstructed from the experimental data, with correction for experimental imperfections. Each color corresponds to the photon number distribution in the output state for the Fock state $\ket m$ at the input. Insets: worst-case fidelities of the reconstructed processes acting within subspaces $\mathcal{H}'_{\hat a,\hat a^\dag}(n)$ of the optical Hilbert space spanned by number states $\ket 1\ldots \ket{n}$ and $\ket 0\ldots \ket{n-1}$, for \hata and \adag, respectively. 
}\label{fig:tensors}
\end{figure}

The diagonal elements $\mathcal{E}^{mm}_{kk}$ of the reconstructed process tensors are shown in Fig.~\ref{fig:tensors}. These elements permit straightforward interpretation: they give the probability that the output of the quantum ``black box" contains $k$ photons when the $m$-photon state is present at the input. According to Eq.~(\ref{an}), we expect $(\mathcal{E}_{\hat a})^{mm}_{kk}= m\delta_{k,m-1}$ and $(\mathcal{E}_{\hat a})^{mm}_{kk}= (m+1)\delta_{k,m+1}$. The experimental result is consistent with this expectation and explicitly features the ``raising" and ``lowering" properties of \adag and $\hat a$. For input state $\ket m$, the output state is similar to $\ket{m+1}$  for operator \adag and $\ket{m-1}$ for operator \hata. The height of the bars in the plots increases linearly with $m$, which is associated with the squared factors $\sqrt{m+1}$ and $\sqrt m$ in the right-hand sides of  Eq.~(\ref{an}). 

The consistency of the estimated process tensors with those theoretically expected can be quantified using the fidelity benchmark. We estimate the worst-case fidelity between normalized states  $\hat a\ket\psi,\hat a^\dag\ket\psi$ and the respective outputs $\mathcal{E}_{\hat a,\hat a^\dag}(\ketbra\psi\psi)$ of the reconstructed processes. The processes are applied to pure states within subspaces $\mathcal{H}'_{\hat a,\hat a^\dag}(n)$ of the optical Hilbert space spanned by number states $\ket 1\ldots \ket{n}$ and $\ket 0\ldots \ket{n-1}$, respectively, with $n$ ranging between $1$ and $n_{\max}$. For each $n$, we employ the genetic algorithm to identify the input state producing the lowest fidelity in $\mathcal{H}'_{\hat a,\hat a^\dag}(n)$ and plot the corresponding fidelity in the insets of Fig.~\ref{fig:tensors}. The fidelities decrease somewhat with increasing $n$ because, for high photon-number states, the overlap with the probe states is low and hence the experimental data do not provide sufficient information about the effect of the process on these states \cite{anis11}.

To summarize, we have experimentally implemented the operations of optical photon creation and annihilation and used the technique of coherent-state quantum-process tomography to explicitly evaluate, for the first time, their process tensors. The reconstructed process tensors exhibit ``raising" and ``lowering" properties of these operations. This is the first experiment in which complete tomography of trace-non-preserving quantum processes has been carried out.


We thank B. Sanders and C. Simon for helpful discussions and A. J. Hendricks, A. Chandra, A. Fedorov and A. S. Prasad for help in the lab. We acknowledge financial support from NSERC and CIFAR.

\section{Supplementary information}
\paragraph{Theoretical background.} Here we explain why the setups shown in Fig.~\ref{fig:setup} approximately implement the action of the photon annihilation and creation operators. The action of a beam splitter on quantum states in the signal and trigger input modes described by annihilation operators $\hat a_s$ and $\hat a_t$ can be modeled as evolution under Hamiltonian \cite{jex95}
\begin{equation}\label{BSHam}
\hat H_{\rm BS}=i\lambda \hat a_s \hat a_t^\dag-i\lambda \hat a_s^\dag \hat a_t,
\end{equation}
where $\lambda$, assumed real, is related to the amplitude reflectivity $r$ of the beam splitter as $r=\sin(\zeta)$, with $\zeta=\lambda\tau/\hbar$ and $\tau$ being the fictitious interaction time. If $|\zeta|\ll 1$, the evolution under Hamiltonian \eqref{BSHam} can be approximated as
\begin{equation}\label{BSevo}
e^{-i\hat H\tau/\hbar}\approx 1-i\hat H\tau/\hbar.
\end{equation}
The beam splitter input consists of the target state $\ket\psi$ in the signal channel and the vacuum state in the trigger channel. The evolution operator, acting on this state, yields
\begin{equation}
e^{-i\hat H_{\rm BS}\tau/\hbar}\left(\ket\psi_s\ket 0_t\right)\approx\ket\psi_s\ket 0_t+\zeta\left(\hat a_s\ket\psi_s\right)\ket 1_t.
\end{equation}
Conditioning on single-photon detection in the trigger channel, we obtain state $\hat a_s\ket\psi_s$ in the signal.

The parametric down-conversion is characterized by the simplified Hamiltonian
\begin{equation}\label{PDCHam}
\hat H_{\rm PDC}=i\lambda \hat a_s^\dag\hat a_t^\dag-i\lambda \hat a_s \hat a_t.
\end{equation}
In this case, $\zeta=\lambda\tau/\hbar$ has the meaning of the squeezing parameter. Applying this Hamiltonian under approximation \eqref{BSevo} to the tensor product of the target and vacuum state, we obtain
\begin{equation}
e^{-i\hat H_{\rm PDC}\tau/\hbar}\left(\ket\psi_s\ket 0_t\right)\approx\ket\psi_s\ket 0_t+\zeta\left(\hat a_s^\dag\ket\psi_s\right)\ket 1_t,
\end{equation}
which projects onto the signal state $\hat a_s^\dag\ket\psi_s$ if a trigger photon is detected.

The above treatment is valid as long as higher-order terms in the Taylor decomposition \eqref{BSevo} can be neglected. This is the case when $\zeta n_{\ket\psi}\ll 1$, with $n_{\ket\psi}$ being the mean photon number in the target state.

\begin{figure}[h]
\includegraphics[width=\columnwidth]{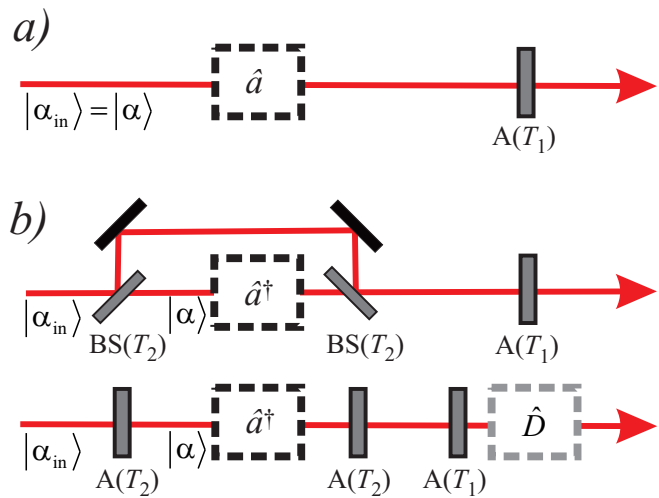}
\caption{Model of experimental imperfections for the photon annihilation (a) and creation (b) setups. The top and bottom schemes in (b) are equivalent to each other. Notation: BS, beam splitter; A, attenuator; the quantities in parentheses denote the transmission of the corresponding optical element. $\hat D$ denotes the operator of phase-space displacement  by $\Delta X=\sqrt {2T_1}(1-T_2)\alpha_{\rm in}$. The amplitude of the input coherent probe state is $\alpha_{\rm in}$ while $\alpha$ is the effective amplitude of the coherent state in the mode acted upon by the operators.}
\label{fig:imperfections}
\end{figure}

\paragraph{Experimental imperfections.} The models used to account for the experimental imperfections are shown in Fig.~\ref{fig:imperfections}. For the process associated with operator $\hat a$, the detrimental effects are the linear losses, non-unitary quantum efficiency and the electronic noise of the homodyne detector \cite{appel07}. All these effects can be quantified and their cumulative contribution modeled by an attenuator with transmission $T_1=0.75$ placed after a ``black box" containing an ideal photon annihilation operator [Fig.~\ref{fig:imperfections}(a)]. Accounting for linear losses in maximum-likelihood homodyne reconstruction is a well-known technique, which consists of modifying the measurement operator associated with detecting field quadrature values \cite{leobook,lvov04,anis11}.

For the photon creation, correcting for the imperfections is more complicated, because one needs to take into account the mode mismatch between the probe field and the signal mode of parametric down-conversion, which is determined by the mode of the pump and the optics in the idler channel \cite{aichele02}. This mismatch is modeled by a Mach-Zehnder  interferometer with beam splitting ratio $T_2$ and an ideal photon creation operator placed into one of its arms [Fig.~\ref{fig:imperfections}(b), top].
To determine $T_2$, we observe that, when the input $\ket{\alpha_{\rm in}}$ is in the vacuum state, the output is expected to be a statistical mixture of the single-photon and vacuum states, with the single-photon fraction equal to $T_1T_2$. The output that we observe experimentally in this case is indeed, to a high degree of precision, described by such a mixture \cite{Fock,sim09}, with the single-photon fraction of $0.59$. Knowing that $T_1=0.75$, we conclude that $T_2=0.59/0.75=0.79$.

With this model, if the coherent state entering the interferometer is $\ket{\alpha_{\rm in}}$, the state entering the ``black box" is $\ket\alpha=\ket{\sqrt{T_2}\alpha_{\rm in}}$. The latter is the amplitude that we refer to  in Figs.~\ref{fig:Wigners} and \ref{fig:countrates}(b) as well as when using \eeqref{Erho} to evaluate the process tensor of the photon creation operator [Fig.~\ref{fig:tensors}(b)]. Note that our model is corroborated by the photon statistics shown in Fig.~\ref{fig:countrates}(b). The fit to the experimental data yields $\pr_{\hat a^\dag}(\alpha)\propto1+0.98\alpha^2$ which is close to the theoretically expected $\pr_{\hat a^\dag}(\alpha)\propto1+\alpha^2$.

For the purpose of accounting for the imperfections in the process reconstruction, it is convenient to reformulate their model in terms of the scheme shown in the bottom panel of Fig.~\ref{fig:imperfections}(b).  The output of the ``black box" undergoes a linear loss channel with transmissivity $T_1T_2$ followed by phase-space displacement by $\Delta X=\sqrt {2T_1}(1-T_2)\alpha_{\rm in}$. These transformations must be compensated for in the process reconstruction. To that end, prior to launching the iterative algorithm with correction for a linear loss, we apply reverse displacement to the measured quadrature data. Specifically, we subtract $\Delta X \cos\theta$ from each experimentally measured sample of quadrature observable $\hat X_\theta=\hat X\cos\theta+\hat P\sin\theta$, where $\theta$ is the local oscillator phase. The modified values are then used as input for the iterative algorithm.

\begin{figure}[h]
\includegraphics[width=\columnwidth]{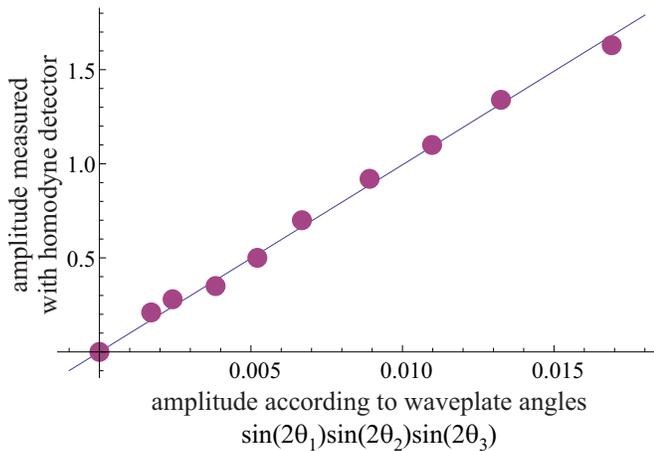}
\caption{Calibration of the coherent state amplitude by homodyne measurement.}
\label{fig:cohcal}
\end{figure}

\paragraph{Calibrating coherent state amplitudes.} The probe coherent states are generated by a series of three variable attenuators, each made up of a half-wave plate and a polarizer, followed by neutral-density attenuators. The amplitude of the coherent state is then given by $\alpha_{\rm in}=A\sin 2\theta_1\sin 2\theta_2\sin 2\theta_3$, where $\theta$'s are the waveplate orientation angles and $A$ is a common factor. Determining $A$ directly from power measurements proved unpractical. Instead, we calibrated this factor by measuring the amplitudes of the coherent states $\ket{\alpha_{\rm out}}$  observed with the homodyne detector in the absence of trigger events in the photon annihilation setup [Fig.~\ref{fig:setup}(a)]. This approach is more reliable because an absolute gauge for the amplitude scale is provided by the quadrature variance of the vacuum state. Once $\alpha_{\rm out}$ has been evaluated (Fig.~\ref{fig:cohcal}), the input coherent state amplitude is determined by taking into account detection losses, i.e. according to $\alpha_{\rm out}=\sqrt{T_1}\alpha_{\rm in}$.

\end{document}